\documentclass[final]{svjour3}
\usepackage[dvipdfmx]{graphicx}
\usepackage{rotating}
\usepackage{amssymb}
\usepackage{mathptmx}
\usepackage[colorinlistoftodos]{todonotes}
\usepackage{array} 
\usepackage{color}
\usepackage{url}
\usepackage{subfig}
\usepackage[numbers]{natbib}
\makeatletter
\journalname{Journal of Low Temperature Physics}

\bibpunct{}{}{,}{s}{}{,}

\begin{document}

\newcommand{\hdblarrow}{H\makebox[0.9ex][l]{$\downdownarrows$}-}
\title{Fabrication of antenna-coupled KID array for Cosmic Microwave Background detection}

\author{Q. Y. Tang$^1$ \and P. S. Barry$^1$ \and R. Basu Thakur$^1$, \and A. Kofman$^2$ \and J. Vieira$^2$ \and E. Shirokoff$^1$} 
\institute{$^1$Kavli Institute of Cosmological Physics, University of Chicago,\\ Chicago, IL 60637, USA\\
$^2$Department of Astronomy, University of Illinois at Urbana-Champaign,\\ Urbana, IL 61801, USA\\
\email{tangq@uchicago.edu}}

\maketitle

\begin{abstract}
Kinetic Inductance Detectors (KIDs) have become an attractive alternative to traditional bolometers in the sub-mm and mm observing community due to their innate frequency multiplexing capabilities and simple lithographic processes. These advantages make KIDs a viable option for the $O(500,000)$ detectors needed for the upcoming Cosmic Microwave Background - Stage 4 (CMB-S4) experiment. We have fabricated antenna-coupled MKID array in the 150GHz band optimized for CMB detection. Our design uses a twin slot antenna coupled to inverted microstrip made from a superconducting Nb/Al bilayer and SiN$_x$, which is then coupled to an Al KID grown on high resistivity Si. We present the fabrication process and measurements of SiN$_x$ microstrip resonators.
%
%
\end{abstract}

\section{Introduction}\label{sec:intro}
In the last few decades, studies of the Cosmic Microwave Background (CMB) have made tremendous contributions to our understanding of the cosmological universe. The measurements are in astounding agreement with a universe of a $\Lambda$CDM cosmology [1], the standard model of Big Bang cosmology. Despite the successes, there still remain many science goals for CMB observers. CMB-Stage 4 (S4) is the next generation CMB experiment and will consist of a collaboration of telescopes located at the South Pole and Atacama plateau in Chile. To reach the target sensitivity, the number of detectors will increase from $O(10,000)$ of current CMB experiments to $O(500,000)$ [2]. Traditionally, CMB telescopes use transition edge sensor (TES) bolometers as detectors. Although they are well-understood and characterized, TES bolometers face several challenges in scaling up for CMB-S4, especially with regards to complicated fabrication process, which requires precise control, and difficulty in multiplexing, due to the readout electronics needed [3,4].

Kinetic inductance detectors (KIDs) are an alternative technology that offers simple designs and intrinsic multiplexing capabilities. With these advantages, KIDs present a feasible solution for the next generation CMB experiments. Several non-CMB millimeter and sub-millimeter wavelength experiments have deployed, or are currently developing, KIDs such as SuperSpec, NIKA, and MAKO [5-7]. Current developments in KID arrays for CMB experiments employ lumped element KIDs (LEKIDs), as seen in [8]. In this paper, we present the fabrication process of an antenna-coupled LEKID array for CMB detection. 

\section{Detector Design and Fabrication}
\subsection{Design}

The current design uses two twin-slot antennas, sensitive to the two orthogonal polarizations. The dimensions of the antenna slots were chosen to have a well-defined beam (half angle $\approx 15^\circ$) at 150GHz. We couple the antenna to an inverted microstrip transmission line. The microstrip consists of a bottom layer of Nb/Al bilayer, which forms the microstrip, a top layer of Nb as the ground plane, and a layer of silicon nitride (SiN$_x$) in between as the dielectric. The microstrip carries the mm-wave signal from the antenna to our detectors, which are Al LEKIDs, comprising a discrete inductive meander and interdigitial capacitor (IDC). The Al KIDs are coupled capacitively to a co-planar waveguide (CPW) transmission line for readout, made in the same bottom Nb layer of the microstrip. For details regarding antenna and detector design, refer to [9]. Once we demonstrate the performance of these prototype devices, we will look to integrate these detectors with a wide-band antenna (e.g. sinuous antenna [3]) for simultaneous coverage of the CMB frequency bands of interest.

\subsection{Fabrication}
\label{sec:fab}
\begin{figure}
  \centering
  \includegraphics[width=0.8\columnwidth]{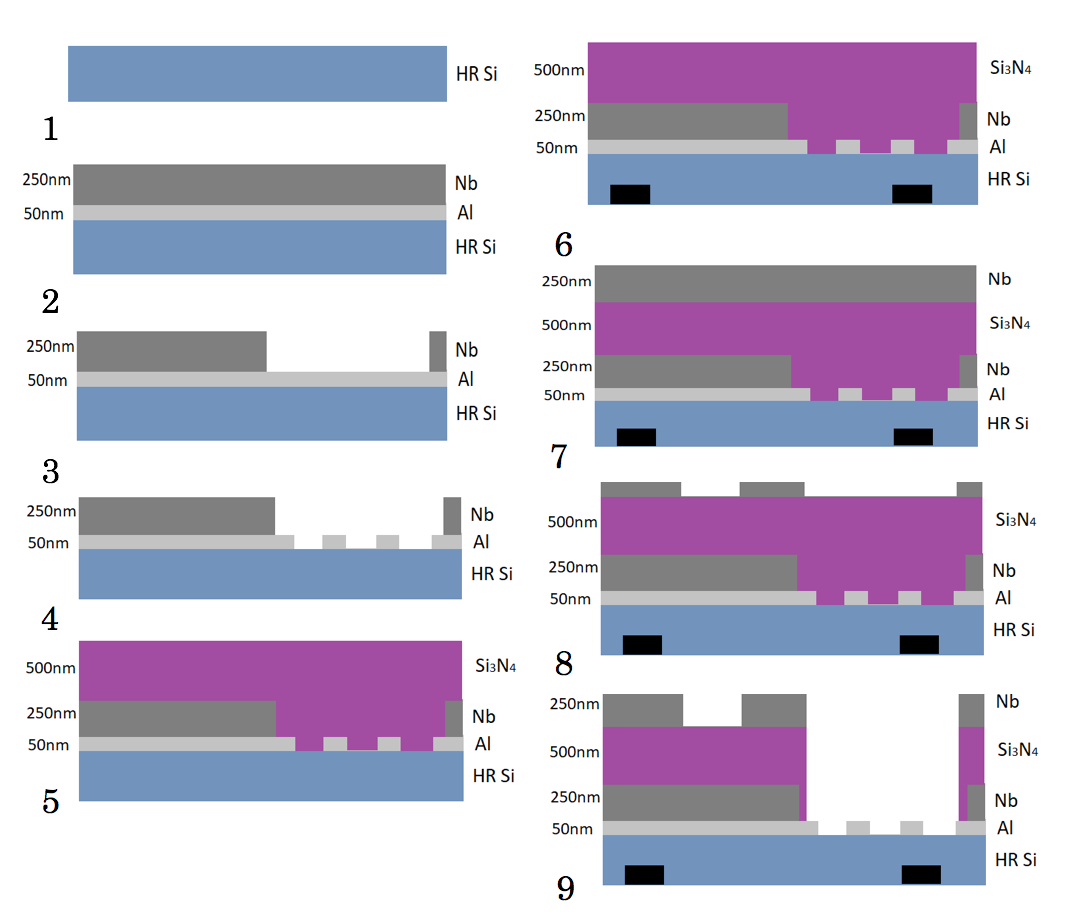}
  \caption{\label{fig:fabdiagram}Cartoon of cross section view of the wafer during all the processing steps. The bottom right figure shows the microwave readout region, the IDC, and the mm-wave transmission line from left to right. See text for details of process steps.}
\end{figure}

Fabrication is done on a high resistivity ($>$ 4k$\Omega$) silicon wafer. Figure~\ref{fig:fabdiagram} shows the process flow in detail and contains the step numbers referred by the text below in parentheses. 

The wafer is first cleaned with a sonicated acetone bath, IPA and water bath. A hydrofluoric (HF) acid bath dip removes native oxide followed by a 150C vacuum bake for 15 minutes to dehydrate the wafer (step 1). We then deposit 50nm of Al, immediately followed by 250nm of Nb using an electron beam (e-beam) evaporator without breaking vacuum (step 2). 
Prior to each lithography step, the wafer is cleaned using acetone, isopropyl alcohol (IPA), and deionized (DI) water baths and then vacuum baked at 90C for 3 minutes to dehydrate the surface. After spinning resist (cf. Table \ref{tab:litho}), we expose the pattern using a maskless lithography writer from Heidelberg Instruments \footnote{https://himt.de/index.php/maskless-write-lasers.html}. 

\begin{figure}
\centering
\subfloat{{\includegraphics[height=2.9cm]{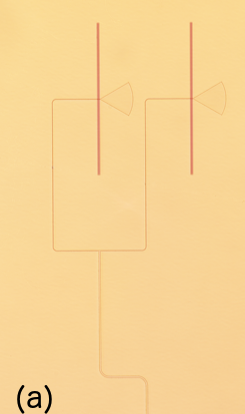}}}
\hfill
\subfloat{{\includegraphics[height=2.9cm]{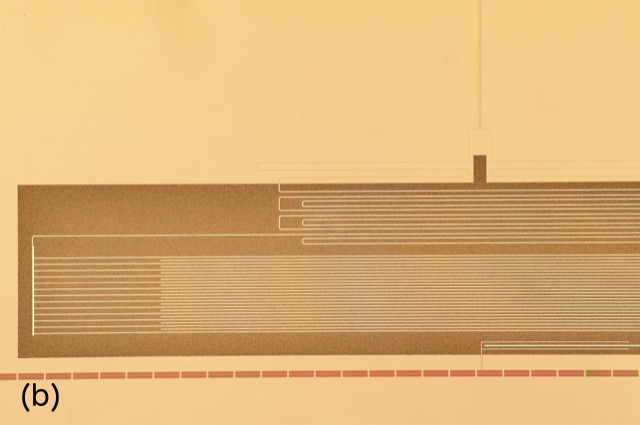}}}
\hfill
\subfloat{{\includegraphics[height=2.9cm]{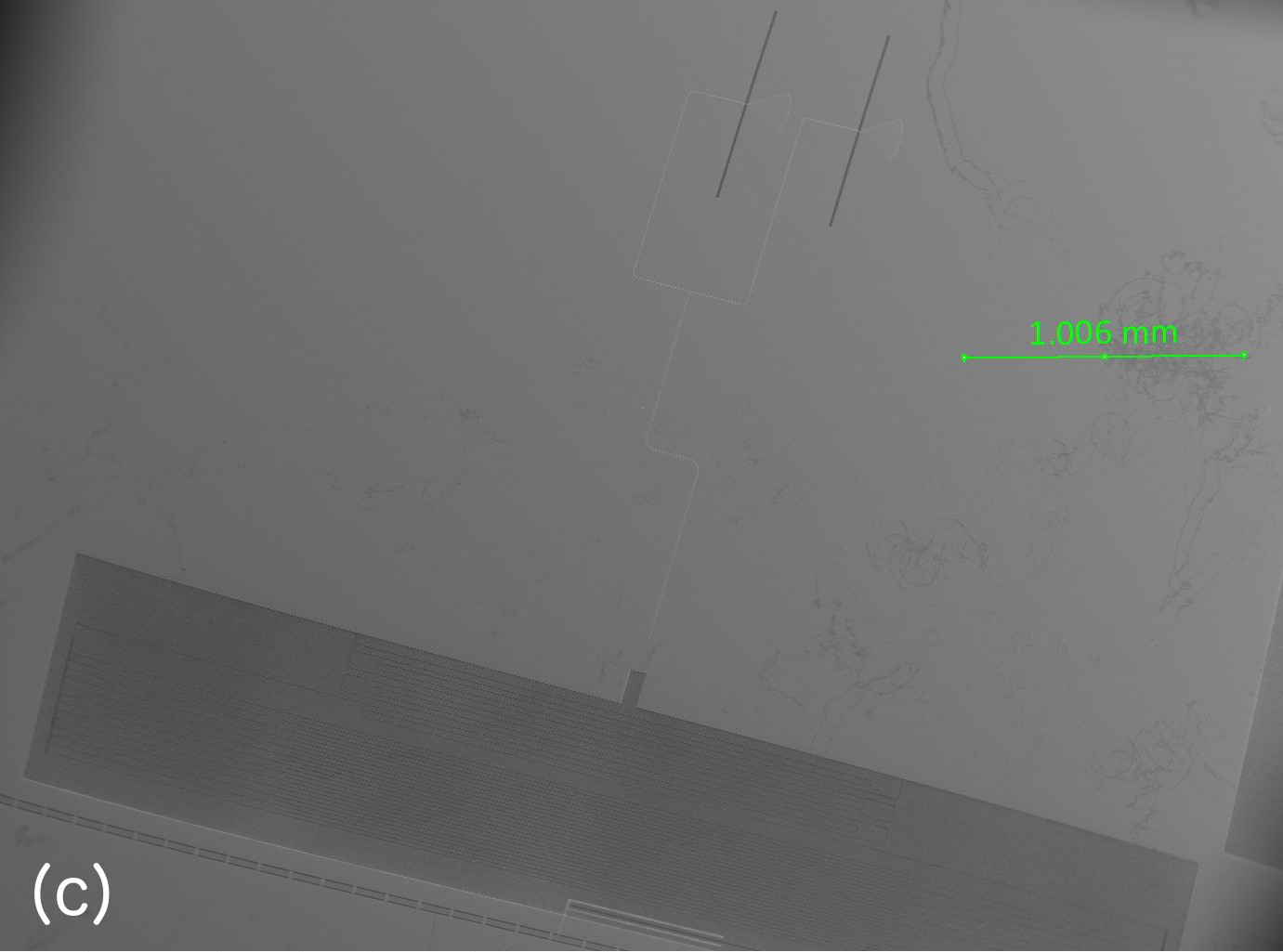}}}
\caption{\label{fig:fabpics} {\it Left}.  Optical microscope image of the antenna and Nb microstrip structure. {\it Middle}. Optical microscope image of the feedline coupled to the KID. The KID is capacitively coupled to the Nb feedline at the bottom. {\it Right}. Scanning electron microscope (SEM) image of a fabricated single polarization antenna-coupled KID.}
\end{figure}

The Nb layer is then dry etched using an inductively coupled plasma (ICP) fluorine etcher (step 3). We then pattern the bottom Al KIDs layer, which is then etched using a standard Al wet etchant containing phosphoric and nitric acid (step 4). To remove the photoresist after etching, we use an O$_2$ plasma asher at 70C and 300W for 240 seconds to descum the wafer before using a NMP (1-Methyl-2-pyrrolidon) bath heated at 80C. This is then followed by ultrasonic NMP bath, IPA, and DI water rinse. We then grow the dielectric layer of our microstrip, 500nm of SiN$_x$, over our wafer (step 5) using High Density Plasma Chemical Vapor Deposition (HPDCVD). After depositing the dielectric layer, we pattern alignment features on the backside of our wafer and etch 80$\mu$m into the silicon wafer with the deep silicon Reactive Ion Etcher (RIE) (cf. Section~\ref{sec:lsw}) (step 6). We then deposit another 250nm of Nb on top of the SiN$_x$ layer which acts as the groundplane (step 7). Finally, we etch the antenna slots (step 8) and remove the SiN$_x$ (step 9) using a fluorine-based ICP etcher to remove the dielectric from the IDCs and CPW wirebond pads. The specific parameters used in each lithography and plasma etch/deposition are shown in Tables~\ref{tab:litho} and~\ref{tab:recipes}, respectively. Microscope images of the final devices are shown in Figure~\ref{fig:fabpics}.


\begin{table}
\centering
\begin{tabular}{c c c c c }
\hline \hline
Step & Layer & Photoresist & Laser (nm) & Dosage \\
& & & & (mJ/cm$^2$) \\
\hline
2,8 & Nb CPW/ground plane & AZ 1512 & 375 & 100 \\
3 & Al KIDs & AZ 1512 & 375 & 130 \\
6& Si holes/lenses &  AZ 1518 & 405 & 135 \\
9& SiN$_x$ & AZ 1512 & 375 & 110 \\ 
\hline
& SU-8 posts & SU-8 3050 & 375 & 245 \\
\hline \hline
\end{tabular}
\caption{\label{tab:litho} Lithography details for fabrication of detector wafer and lens seating wafer. Step number matches the process step shown in Figure~\ref{fig:fabdiagram}. }

\end{table}
\begin{table*}[tb!]
\centering
\setlength\tabcolsep{4pt} 
\begin{tabular}{c c c c c c c c c  } 
\hline \hline
Step & Recipe & ICP Power & Bias Power & Pres. & He Pres. &  Gas 1 flow & Gas 2 flow & Gas 3 flow \\
 &  & (W) & (W) &(mTorr) & (Torr) & (sccm) & (sccm) &  (sccm)\\
\hline
2,8 & Nb etch & 600 & 	50&	50&	5	& CF$_4$, 40& CHF$_3$, 10 & Ar,	10  \\
5 & SiN$_x$ dep& 400 &20 & 10& 5 & SiH$_4$, 20 & N$_2$, 2 & Ar, 20\\
9 & SiN$_x$ etch & 100 &	20&	30&	5&	SF$_6$, 20 & -&- \\
\hline \hline

\end{tabular}
\caption{\label{tab:recipes} Recipes used during fabrication steps. Step number matches the process step shown in Figure~\ref{fig:fabdiagram}.}
\end{table*}

\subsection{Lens Seating Wafers} \label{sec:lsw}

\begin{figure}[h]
  \centering
\subfloat{{\label{fig:lenses}\includegraphics[height=2.9cm]{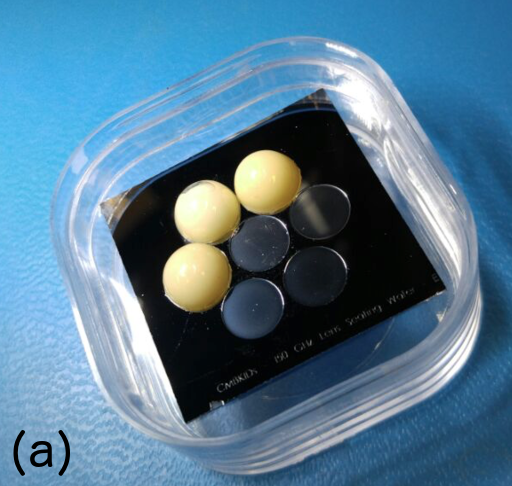}}}
\hspace{4em}\subfloat{{\label{fig:alignnotalign}\includegraphics[height=2.9cm]{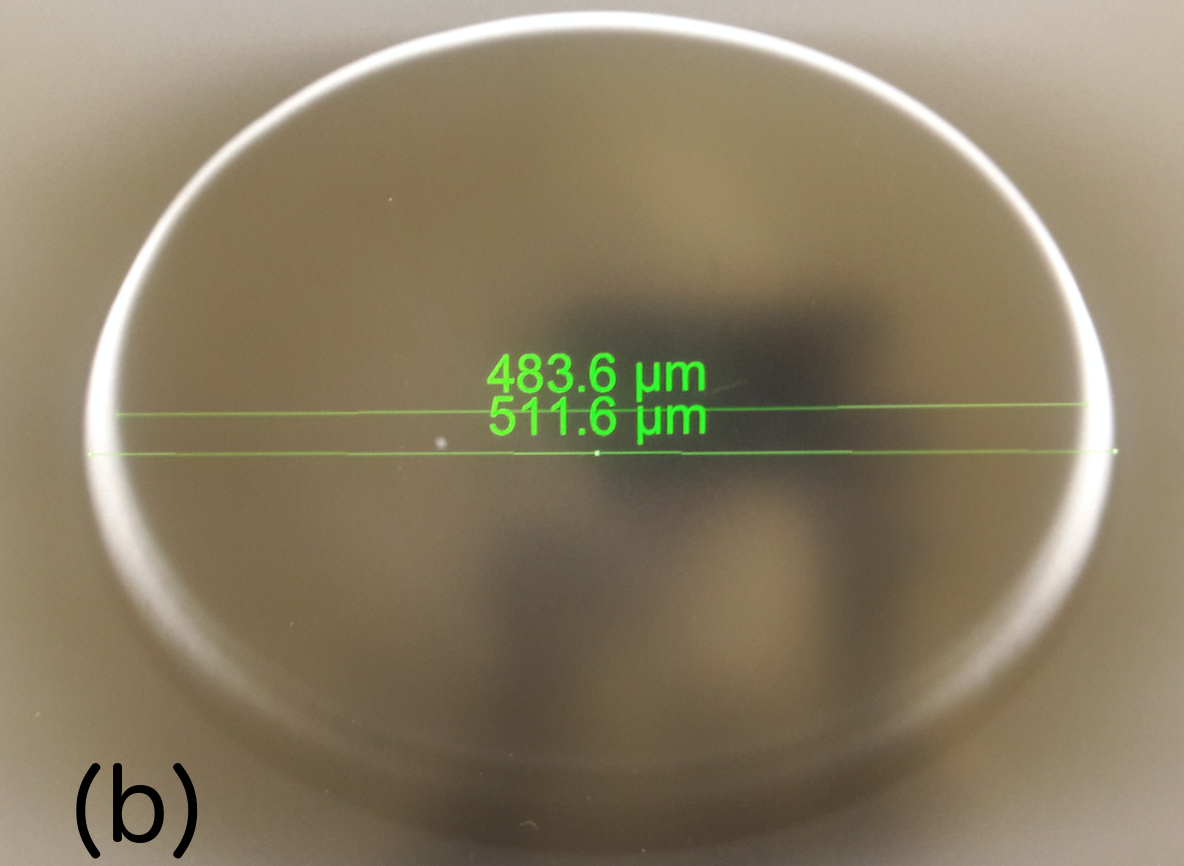}}}

  \caption{{\it Left}. Photograph of the lens seating wafer partially populated with alumina lenslets. {\it Right}. SEM image of a SU-8 post on the lens seating wafer. The diameter is measured to be roughly 500$\mu$m and 70$\mu$m tall. }
\end{figure}

To focus radiation onto the antenna, we use 1/4-inch diameter alumina lenslets with a thermoformed polyetherimide anti-reflection (AR) coating optimized for 150GHz. The spherical lenses are placed into a seating wafer that provides a well-defined extension length to maximize the antenna directivity [10]. An image of the lens wafer with lenses is shown in Figure~\ref{fig:lenses}.


\begin{figure}[h]
  \centering
\subfloat{{\label{fig:alignment}  \includegraphics[height=2.7cm]{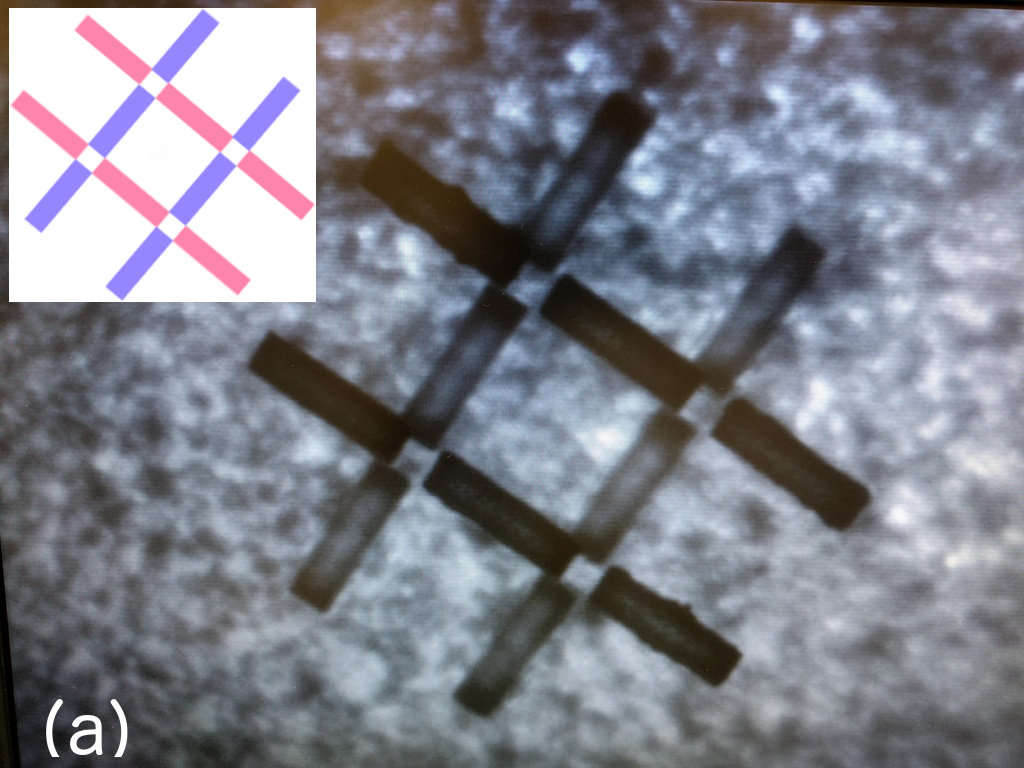}}}
\hfill\subfloat{{\label{fig:alignnotalign}  \includegraphics[height=2.7cm]{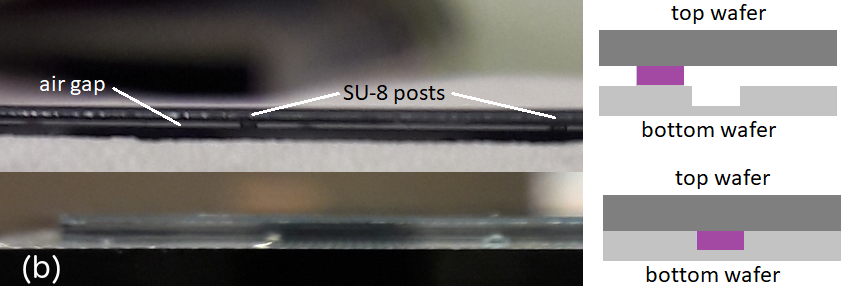}}}
\caption{{\it Left}. IR microscope image of the alignment of SU-8 posts to holes by viewing two wafers with different features lying one on top of another. Each rectangular bar measures 50$\mu$m x 200$\mu$m. Rectangles of the same orientation are on the same wafer. When perfectly aligned, the ends of four orthogonal rectangles should form a square, as shown in the upper left inset image. We estimate misalignment to be no more than $\approx$20$\mu$m. {\it Right}. Side view of the lens wafer when the posts are misaligned with the holes (top) and aligned (bottom). A cartoon is shown as a comparison to illustrate SU-8 post and hole alignment. The posts are shown in purple and wafers in gray.}
\end{figure}

The lenslets are glued onto the lens seating wafer, which is made from a 500$\mu$m thick double-sided polished silicon substrate. We pattern the lenslet holes, as described in Table~\ref{tab:litho}, and subsequently etch 250$\mu$m into the wafer. Alignment between the antenna and lenses is critical to have a well-defined beam. To do this, we pattern posts made from SU-8 3050, a permanent epoxy negative photoresist, on the backside of the wafer, with parameters specified in Table~\ref{tab:litho}. After the photoresist is developed, the wafer is baked at 160C for 2 hours to cure the SU-8 posts. Each post is approximately $ 70\mu$m tall and 500$\mu$m wide in diameter, designed to fit the 80$\mu$m deep holes on the backside of the detector wafer. To measure the misalignment, Figure~\ref{fig:alignment} shows the mismatch between two wafers aligned this way viewed through an infrared (IR) camera microscope. We observe a maximal misalignment of $\approx$20 $\mu$m after multiple realignments of the wafer assembly. Figure~\ref{fig:alignnotalign} shows the side view of the lens wafer when the holes and posts are aligned and misaligned. 

\section{Microstrip Results and Conclusion}

To measure the loss in the microstrip carrying the microwave signal from antennae to the KID, we design and fabricate microstrip resonators, which have a resonance frequency of $\sim 5$GHz, using the same materials as described in Section~\ref{sec:fab}. The ground plane with the CPW lines is made by e-beam deposition of 175nm of Nb on high resistivity silicon wafers ($>$4k$\Omega$). 500nm of SiN$_x$ is then grown on top, followed by a deposition of 225nm of Nb sputtered using lift-off to form the microstrip resonators. The lithography, SiN$_x$ deposition and etch recipes used are the same as the ones in 
Table~\ref{tab:litho} and~\ref{tab:recipes}. 



\subsection{Experimental Setup}
We test our resonators in a Model 104 Olympus cryostat made by High Precision Devices, Inc. Two internal stages are mounted on the 4K plate, for which the refrigerants are used to reach temperatures of $\approx$ 500mK and 50mK (the detector stage). This stage has a temperature range of 50mK to $\approx$ 2K. We connect the device output line to CIT-CRYO-12A high electron mobility transistor amplifier mounted on the 4K stage, which has a gain of $\sim$20dB from 1-12GHz. DC blocks are placed on both the input and output lines on the 50K, 4K, and 500mK stages for thermal isolation and two cryogenic 20dB attenuators are placed along the input to the device to minimize thermal noise. The total attenuation in the readout lines is $\sim $70dB.

\subsection{Data Fitting}
We test our devices by mapping the complex transmission as a function of frequency for each resonator, by sending tones from the vector network analyzer to obtain the two-port $S_{21}$ transmission function. The resonance can be modeled as a Lorentzian parametrized mainly by $f_r$, the resonance frequency, and $Q_r$, the total quality factor which describes the width of the resonance. $f_r$ is determined by the inductance and capacitance of the KID. $Q_r$ can be decomposed into $Q_c$, a coupling quality factor describing the coupling capacitance of the KID to the transmission line, and $Q_i$, the intrinsic quality factor of the KID. The relation is defined by $Q_r^{-1} = Q_c^{-1} + Q_i^{-1}$. We fit to the $S_{21}$ data curve with 
\begin{equation} \label{eq:S21}
 S_{21}(x)=\frac{1/Q_i +2i \left(x+\delta f / f_r \right) } {1/Q_r + 2ix}
\end{equation}
where $x = (f-(f_r + \delta f))/f_r + \delta f$, and $\delta f$ is an asymmetry term accounting for any impedance mismatch and standing waves in the circuitry. We use a Markov Chain Monte Carlo formalism to fit the model to our data, as described by [11,12].

\subsection{Results}

\begin{figure}[ht]
  \centering
  \includegraphics[width=0.5\columnwidth]{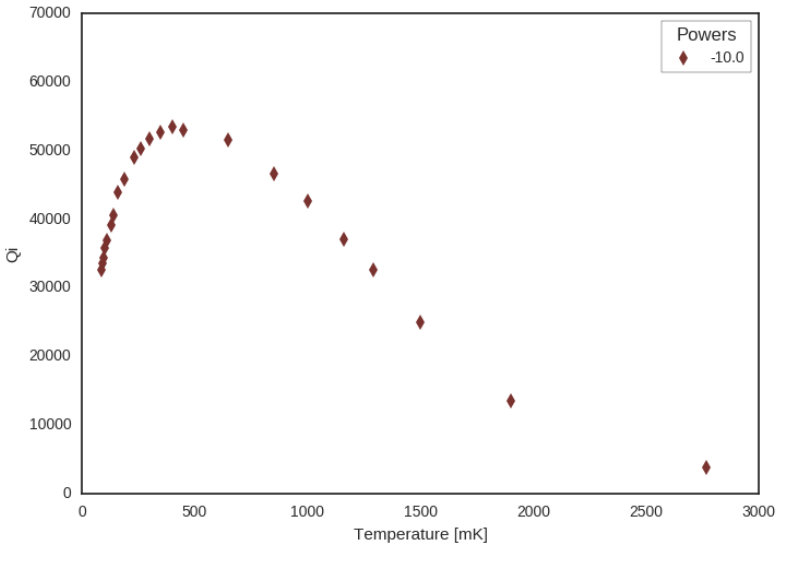}
  \caption{\label{fig:microstrip} Plot of $Q_i$ values obtained from S21 curves at different temperature of the Nb microstrip resonator.}
\end{figure}

Our measurements demonstrate the feasibility of fabricating a low-loss microstrip using SiN$_x$. Figure~\ref{fig:microstrip} shows the $Q_i$ values as a function of stage temperature, inferred from the $S_{21}$ curve fitting at each temperature using Eq. (\ref{eq:S21}). The ambient bath temperature on the focal plane of a CMB telescope is estimated to be between 200-300mK. We measure $Q_i\approx 40,000$ at this temperature and $f_r \sim 5$GHz, sufficiently low-loss ($\tan \delta = 1/Q \approx 2.5\times 10^{-5}$).


In conclusion, we have fabricated a prototype antenna-coupled KID array intending for CMB applications. Additionally, we have measured the SiN$_x$ microstrip structure to be very low-loss at 5GHz, with $Q_i \approx 40,000$ at $\sim$200mK. For performances of our fabricated Al CMB KID devices, which have been measured $Q_i \approx 10^6$, refer to [13]. We are currently exploring other materials which have smaller band gap energies, such as titanium nitride and aluminum manganese, to implement the 90GHz detection band. 

\begin{acknowledgements}
This work is partially supported by NSF award \#1554565 and the Kavli NSF-PFC3 Detector Development grant. This work was supported in part by the Kavli Institute for Cosmological Physics at the University of Chicago through
grant NSF PHY-1125897 and an endowment from the Kavli Foundation and its founder Fred Kavli. This work made use of the Pritzker Nanofabrication Facility of the Institute for Molecular Engineering at the University of Chicago, which receives support from SHyNE, a node of the National Science Foundation’s National Nanotechnology Coordinated Infrastructure (NSF NNCI-1542205). \end{acknowledgements}


\end{document}